\PassOptionsToPackage{unicode}{hyperref}
\PassOptionsToPackage{hyphens}{url}
\PassOptionsToPackage{dvipsnames,svgnames,x11names}{xcolor}
\documentclass[
  12pt]{article}

\usepackage{amsmath,amssymb}
\usepackage{iftex}
\ifPDFTeX
  \usepackage[T1]{fontenc}
  \usepackage[utf8]{inputenc}
  \usepackage{textcomp} % provide euro and other symbols
\else % if luatex or xetex
  \usepackage{unicode-math}
  \defaultfontfeatures{Scale=MatchLowercase}
  \defaultfontfeatures[\rmfamily]{Ligatures=TeX,Scale=1}
\fi
\usepackage{lmodern}
\ifPDFTeX\else  
\fi
\IfFileExists{upquote.sty}{\usepackage{upquote}}{}
\IfFileExists{microtype.sty}{% use microtype if available
  \usepackage[]{microtype}
  \UseMicrotypeSet[protrusion]{basicmath} % disable protrusion for tt fonts
}{}
\makeatletter
\@ifundefined{KOMAClassName}{% if non-KOMA class
  \IfFileExists{parskip.sty}{%
    \usepackage{parskip}
  }{% else
    \setlength{\parindent}{0pt}
    \setlength{\parskip}{6pt plus 2pt minus 1pt}}
}{% if KOMA class
  \KOMAoptions{parskip=half}}
\makeatother
\usepackage{xcolor}
\makeatletter
\ifx\paragraph\undefined\else
  \let\oldparagraph\paragraph
  \renewcommand{\paragraph}{
    \@ifstar
      \xxxParagraphStar
      \xxxParagraphNoStar
  }
  \newcommand{\xxxParagraphStar}[1]{\oldparagraph*{#1}\mbox{}}
  \newcommand{\xxxParagraphNoStar}[1]{\oldparagraph{#1}\mbox{}}
\fi
\ifx\subparagraph\undefined\else
  \let\oldsubparagraph\subparagraph
  \renewcommand{\subparagraph}{
    \@ifstar
      \xxxSubParagraphStar
      \xxxSubParagraphNoStar
  }
  \newcommand{\xxxSubParagraphStar}[1]{\oldsubparagraph*{#1}\mbox{}}
  \newcommand{\xxxSubParagraphNoStar}[1]{\oldsubparagraph{#1}\mbox{}}
\fi
\makeatother

\usepackage{longtable,booktabs,array}
\usepackage{calc} % for calculating minipage widths
\usepackage{etoolbox}
\makeatletter
\patchcmd\longtable{\par}{\if@noskipsec\mbox{}\fi\par}{}{}
\makeatother
\IfFileExists{footnotehyper.sty}{\usepackage{footnotehyper}}{\usepackage{footnote}}
\makesavenoteenv{longtable}
\usepackage{graphicx}
\makeatletter
\def\maxwidth{\ifdim\Gin@nat@width>\linewidth\linewidth\else\Gin@nat@width\fi}
\def\maxheight{\ifdim\Gin@nat@height>\textheight\textheight\else\Gin@nat@height\fi}
\makeatother
\setkeys{Gin}{width=\maxwidth,height=\maxheight,keepaspectratio}
\makeatletter
\def\fps@figure{htbp}
\makeatother

\makeatletter
\@ifpackageloaded{caption}{}{\usepackage{caption}}
\AtBeginDocument{%
\ifdefined\contentsname
  \renewcommand*\contentsname{Table of contents}
\else
  \newcommand\contentsname{Table of contents}
\fi
\ifdefined\listfigurename
  \renewcommand*\listfigurename{List of Figures}
\else
  \newcommand\listfigurename{List of Figures}
\fi
\ifdefined\listtablename
  \renewcommand*\listtablename{List of Tables}
\else
  \newcommand\listtablename{List of Tables}
\fi
\ifdefined\figurename
  \renewcommand*\figurename{Figure}
\else
  \newcommand\figurename{Figure}
\fi
\ifdefined\tablename
  \renewcommand*\tablename{Table}
\else
  \newcommand\tablename{Table}
\fi
}
\@ifpackageloaded{float}{}{\usepackage{float}}
\floatstyle{ruled}
\@ifundefined{c@chapter}{\newfloat{codelisting}{h}{lop}}{\newfloat{codelisting}{h}{lop}[chapter]}
\floatname{codelisting}{Listing}

\makeatother
\makeatletter
\@ifpackageloaded{caption}{}{\usepackage{caption}}
\@ifpackageloaded{subcaption}{}{\usepackage{subcaption}}
\makeatother

\ifLuaTeX
  \usepackage{selnolig}  % disable illegal ligatures
\fi
\usepackage[]{natbib}
\usepackage{bookmark}

\IfFileExists{xurl.sty}{\usepackage{xurl}}{} % add URL line breaks if available
\hypersetup{
  pdftitle={Title},
  pdfauthor={Author 1; Author 2},
  pdfkeywords={3 to 6 keywords, that do not appear in the title},
  colorlinks=true,
  linkcolor={blue},
  filecolor={Maroon},
  citecolor={Blue},
  urlcolor={Blue},
  pdfcreator={LaTeX via pandoc}}

\newcommand{\anon}{1}

\usepackage{enumitem}
\usepackage{booktabs}
\usepackage{cancel}
\usepackage{amssymb,amsmath}
\usepackage{graphics}
\usepackage{color}
\usepackage{xcolor}
\usepackage{soul}
\usepackage{mathrsfs}
\usepackage{multirow}
\usepackage{hhline}
\usepackage{caption}
\usepackage{subcaption}
\usepackage{graphicx}
\usepackage{amsmath}
\usepackage[capitalise,nameinlink]{cleveref}
\usepackage{lipsum}
\usepackage{amsthm}
\usepackage{float}
\usepackage{url}
\usepackage{tikz}

\usepackage{xr}
\makeatletter
\newcommand*{\addFileDependency}[1]{% argument=file name and extension
\typeout{(#1)}% latexmk will find this if $recorder=0
\@addtofilelist{#1}
\IfFileExists{#1}{}{\typeout{No file #1.}}
}\makeatother

\newcommand{\mb}[1]{\mathbf{#1}}

\mathchardef\mhyphen="2D

\newcommand{\sN}{\mathcal{N}}

\newcommand{\vg}{\mathbf{g}}

\newcommand{\veta}{\boldsymbol{\eta}}

\newcommand{\vzeta}{\boldsymbol{\zeta}}
\newcommand{\vpi}{\boldsymbol{\pi}}
\newcommand{\vtau}{\boldsymbol{\tau}}
\newcommand{\mH}{\mathbf{H}}
\newcommand{\mI}{\mathbf{I}}

\newcommand{\R}{\mathbb{R}}
\newcommand{\E}{\mathbb{E}}

\begin{document}

\def\spacingset#1{\renewcommand{\baselinestretch}%
{#1}\small\normalsize} \spacingset{1}

%%%%%%%%%%%%%%%%%%%%%%%%%%%%%%%%%%%%%%%%%%%%%%%%%%%%%%%%%%%%%%%%%%%%%%%%%%%%%%

\if1\anon
{
  \title{\bf Bayesian ACCESS for Understanding Latent Epidemic Trajectories from Publicly Released Suppressed Data: Application to U.S. Opioid-related Overdose Mortality}
\author{Jiahao Cao, Kehe Zhang, Cici Bauer\thanks{
    %The authors gratefully acknowledge \textit{please remember to list all relevant funding sources in the version that gives all author information}
    Corresponding author}\hspace{.2cm}\\
    Department of Biostatistics and Data Science\\
    Center of Spatial-Temporal Modeling of Applications in Population Sciences\\
    The University of Texas Health Science Center at Houston}
  \maketitle
} \fi

\if0\anon
{
  \bigskip
  \bigskip
  \bigskip
  \begin{center}
    {\LARGE\bf Bayesian ACCESS for Understanding Latent Epidemic Trajectories from Publicly Released Suppressed Data: Application to U.S. Opioid-related Overdose Mortality}
\end{center}
  \medskip
} \fi

% \bigskip
\begin{abstract}
Publicly released health statistics play a central role in characterizing temporal trends and identifying structural changes in population health. However, disclosure limitation through suppression of small cell counts, as implemented in systems such as the Centers for Disease Control and Prevention Wide-ranging ONline Data for Epidemiologic Research (CDC WONDER), produces partially observed count data that complicate statistical inference. These challenges are particularly acute for rare outcomes and subgroup analyses, where suppression is widespread and varies across geographic regions, demographic populations, and time. We propose Bayesian ACCESS (Autoregressive Change-point and Clustering Estimation for Suppressed Count Series), a Bayesian hierarchical framework for inference on latent epidemic trajectories and their structural changes from disclosure-limited health statistics. The proposed model directly represents suppressed count data through a suppression-aware observation model, jointly infers multiple temporal change points and latent trajectories, and borrows information across related geographic and demographic populations while preserving meaningful heterogeneity. We apply Bayesian ACCESS to opioid-related overdose mortality data from CDC WONDER for U.S. states from 1999 to 2024. The analysis identifies distinct subgroup-specific epidemic trajectories and structural changes that would be difficult to characterize using publicly released health statistics without explicitly accounting for data suppression and efficiently borrowing information across subgroups.
\end{abstract}

\noindent%
\vfill

\newpage
\spacingset{1.8} % DON'T change the spacing!

\section{Introduction}\label{sec:intro}

Recent national surveillance data suggest that opioid-related overdose (OOD) mortality in the United States has entered a sustained period of decline following decades of rapid increase, marking a potentially important shift in the trajectory of the opioid epidemic \citep{garnett2026drug}. However, the timing, magnitude, and even existence of this decline remain inferential questions rather than simple descriptive facts. Recent epidemiologic studies have therefore emphasized careful characterization of temporal changes while noting substantial heterogeneity across geographic regions, demographic groups, and substance types rather than assuming a uniform national trend \citep{post2025decline,friedman2026us}. Addressing these questions becomes considerably more difficult at the state, county, or population-subgroup level, where publicly available mortality data are routinely suppressed to protect confidentiality, resulting in sparse and partially observed time series. Consequently, identifying structural changes in OOD mortality temporal trend requires statistical methods that can simultaneously account for disclosure-induced suppression, multilevel heterogeneity, and complex temporal dynamics. More broadly, the scientific objective is not simply to recover suppressed observations, but to make reliable inference on the latent epidemic trajectory, including the structural changes that mark major transitions in its evolution.

The statistical challenges begin with the way public health statistics data are produced and released. To protect individual confidentiality, agencies routinely apply disclosure limitation procedures before disseminating tabulated statistics. Common approaches include aggregation, noise infusion, and suppression of small cell counts \citep{abowd2018us}. Among these, small-cell suppression remains one of the most widely used disclosure limitation strategies and is employed by major federal statistical systems, including the CDC Wide-ranging ONline Data for Epidemiologic Research (WONDER) database \citep{tiwari:impact:2014} and the American Community Survey \citep{freiman2017data,li2023impacts}. In practice, analyses based on these publicly released data often begin by recovering suppressed counts so that standard statistical methods and software can be applied. For the suppression mechanism used in CDC WONDER, \citet{tiwari:impact:2014} proposed substitution-based approaches that leverage population counts and regional mortality rates to recover suppressed mortality counts. In OOD research, \citet{erdman2021} developed a three-step imputation procedure that incorporates longitudinal and demographic information to recover suppressed counts in Massachusetts Prescription Monitoring Program data. These studies demonstrate the importance of accounting for suppression, as ignoring suppressed observations can lead to biased estimation and misleading conclusions. However, deterministic substitution methods that replace suppressed values with single estimates fail to propagate uncertainty arising from the unobserved counts. Multiple imputation \citep{rubin1978multiple,murray2018multiple} provides a principled alternative by generating multiple completed datasets that reflect uncertainty in the imputed values. Closely related to Bayesian inference for incomplete data, multiple imputation has become a standard tool in survey statistics and other applied settings. For example, \citet{quick2019estimating} developed a Bayesian spatial model for imputing suppressed county-level stroke and heart disease mortality data from CDC WONDER, demonstrating the benefits of spatial borrowing of information. More recently, \citet{liu2025multiple} proposed a multiple-imputation approach for longitudinal school enrollment data that accommodates nonlinear relationships through auxiliary information. Despite these advances, existing methods primarily regard suppression as a data reconstruction problem. In many public health applications, however, the scientific objective is not to recover the suppressed counts themselves but to infer the latent spatiotemporal dynamics of the underlying disease process.

A central inferential goal in studying OOD mortality is to identify structural changes in temporal trends that mark important transitions in the course of the epidemic. Statistical modeling of these trajectories is challenging because the opioid epidemic exhibits multiple waves, nonlinear and nonstationary temporal dynamics, together with substantial spatial and sociodemographic heterogeneity, reflecting differences in drug supply, prevention efforts, treatment access, and socioeconomic conditions \citep{soelberg2017us,ciccarone2019triple,lee2021systematic,friedman2026charting}. These complexities have motivated a broad range of epidemiologic and systems modeling approaches. For example, \citet{singh2019opioid} applied log-linear regression to OOD mortality data from 1999--2017 and documented substantial geographic and racial disparities associated with social determinants of health. More recently, \citet{lim2022modeling} developed SOURCE, a dynamic systems model that represents feedbacks and delays among multiple components of the opioid crisis to support policy planning. While these approaches have yielded important epidemiologic insights, they either impose relatively restrictive temporal structures or rely on analyses aggregated across populations, limiting their ability to characterize structural changes in heterogeneous state- and race-specific mortality trajectories. Moreover, much of the existing literature lacks a unified statistical framework for modeling the temporal dynamics of OOD under multilevel heterogeneity while identifying structural changes with principled uncertainty quantification.

Statistical methods for change-point detection have a long history in both the frequentist and Bayesian literature and remain an active area of research \citep{aminikhanghahi2017survey,cappello2023bayesian,li2025change}. In the frequentist literature, change points are typically estimated through optimization-based procedures, with inference often relying on test statistics and asymptotic distributions. A widely used example in epidemiology is joinpoint regression \citep{kim2000permutation}, also known as segmented or piecewise regression, which identifies significant changes in temporal trends and has been routinely applied in cancer surveillance through the National Cancer Institute's Joinpoint Regression Program. In contrast, Bayesian change-point models provide a principled  framework for incorporating prior information and quantifying uncertainty in both the number and locations of change points. Since the early work of \citet{chernoff1964estimating}, Bayesian change-point models have been extended through a variety of model formulations and prior specifications \citep{chib1998estimation,koop2007estimation,ko2015dirichlet}. More recently, change-point detection for multivariate time series has received increasing attention \citep[e.g.,][]{zhang2010detecting,niu2016multiple,dass2015clustering,fan2017empirical}. By exploiting dependence across related time series, multivariate change-point models can borrow information across series, thereby improving statistical robustness and efficiency.

Despite these advances, existing approaches address different components of the problem, including data suppression, change-point detection, flexible temporal modeling, or multivariate dependence, and seldom integrate these features within a unified inferential framework. In particular, existing methods differ substantially in their assumptions regarding temporal dynamics within segments, which directly affect both latent mortality-rate estimation and change-point identification. In epidemiology and public health, temporal trends are commonly modeled under piecewise linear structures on either the original or log-transformed rate scale. For example, \citet{dass2015clustering} modeled observed log lung cancer mortality rates for U.S. states using piecewise linear trends with normal errors, and employed a Dirichlet process approach to cluster states with heterogeneous temporal patterns. \citet{hepler2021multivariate} studied county-level overdose mortality rates in Ohio by modeling the logit of the binomial rate as piecewise linear functions with a single change point, while incorporating spatial dependence through a conditional autoregressive model. In contrast, much of the statistical change-point literature assumes observations within each segment are independent and identically distributed, thereby simplifying the temporal dependence structure. Both modeling strategies can be restrictive when the underlying process exhibits complex nonlinear dynamics, and neither naturally accommodates suppression of small counts in publicly released health statistics. For drug overdose mortality, where periods of rapid exponential growth have been followed by stabilization and more recent declines \citep{jalal2018changing,friedman2026charting}, inference on both latent mortality trajectories and the timing of structural changes can be highly sensitive to these modeling assumptions. 

To address these challenges, we develop Bayesian ACCESS, a hierarchical framework for inference on latent mortality trajectories from publicly released health statistics subject to disclosure-induced suppression. Our approach directly models the count data, linking released observations to latent rates through the known suppression rule, and captures temporal dependence through an autoregressive structure motivated by count time-series and epidemic surveillance models \citep{ferland2006integer,held2005statistical}. To capture structural changes over time, the autoregressive parameters are modeled as piecewise constant, with both the number and locations of change points inferred from the posterior distribution. Geographic and demographic heterogeneity are accommodated by decomposing these parameters into demographic-specific multipliers and state-specific components, with the latter clustered under a mixture-of-finite-mixtures prior \citep{miller2018mixture}.  More broadly, Bayesian ACCESS builds upon the rich literature on multivariate spatiotemporal modeling in public health, including stochastic process models \citep{cressie2011statistics,diggle2013statistical}, vector autoregressive models \citep{stock2001vector}, and dynamic systems models \citep{lim2022modeling,huang2025infectious}. Relative to existing approaches, Bayesian ACCESS has several desirable features: i) Our approach jointly represents suppressed counts, heterogeneous temporal dynamics, and multiple structural changes within a unified probabilistic framework, allowing uncertainty arising from suppressed observations to be propagated directly into inference on temporal trajectories and change-point locations, rather than relying on deletion or imputation; ii) We model temporal dynamics in the latent trajectories through an identity-link autoregressive specification, avoiding rate transformations and yielding interpretable parameters for innovation, autoregressive dependence, and demographic heterogeneity; and iii) We model the temporal dynamics of multiple count series jointly, enabling structured information sharing among trajectories. This stabilizes inference under substantial data suppression while preserving flexibility to capture geographic and demographic heterogeneity.

The remainder of the paper is organized as follows. Section~\ref{sec:data} presents the CDC WONDER OOD mortality data from 1999 to 2024, and illustrates the extent and impact of data suppression on identifying structural changes. Section~\ref{sec:method} introduces the proposed modeling framework, including the model formulation, prior specification, and posterior inference. Throughout, we emphasize modeling choices that improve flexibility and interpretability, and prior formulations that are scientifically meaningful within the context of OOD mortality. Section~\ref{sec:simulation} evaluates the numerical performance of the proposed methods using semi-synthetic simulation data. Section~\ref{sec:real_data_analysis} presents a comprehensive analysis of U.S. OOD mortality using CDC WONDER data, identifying multiple structural changes in temporal trends, and characterizing how the timing and evolution of the opioid epidemic differ across states and racial groups. By explicitly accounting for disclosure-induced suppression and efficiently borrowing information across related trajectories, the proposed framework reveals subgroup-specific epidemic trajectories and change points that would otherwise be difficult, or in many cases impossible, to identify from publicly released health statistics alone. Section~\ref{sec:discussion} concludes with a discussion of the main findings, limitations, and future directions. Technical details and additional numerical studies are provided in the Supplementary Materials.

\section{CDC WONDER OOD Mortality Data}\label{sec:data}

The CDC WONDER database has been a primary data source for understanding mortality trends and has been widely used in studies examining temporal trajectories and changes over time. We extracted annual OOD death counts from CDC WONDER by year and racial group at both the national and state levels, along with corresponding U.S. Census population estimates. We first identified drug poisoning deaths using the underlying cause-of-death codes from the International Classification of Diseases, 10th Revision (ICD-10): X40-X44 (Accidental), X60-X64 (Intentional), X85 (Assault by drugs), and Y10-Y14 (Undetermined). Among these deaths, opioid involvement was identified using multiple cause-of-death codes T40.0 (Opium), T40.1 (Heroin), T40.2 (Other opioids), T40.3 (Methadone), T40.4 (Other synthetic narcotics), and T40.6 (Other and unspecified narcotics). Race categories were harmonized across the 1999--2024 study period to ensure consistent race-specific OOD time series, resulting in four racial groups: White, Black or African American (Black), American Indian or Alaska Native (AIAN), and Asian or Pacific Islander (Asian/PI). Counts between 1 and 9 were suppressed. Crude mortality rates were calculated per 100,000 population, where counts fewer than 20 were flagged as unreliable. Detailed procedures are described in Section~S.1.1 of the Supplementary Materials.

\begin{figure}[ht!]
    \centering
    \includegraphics[width=0.8\linewidth]{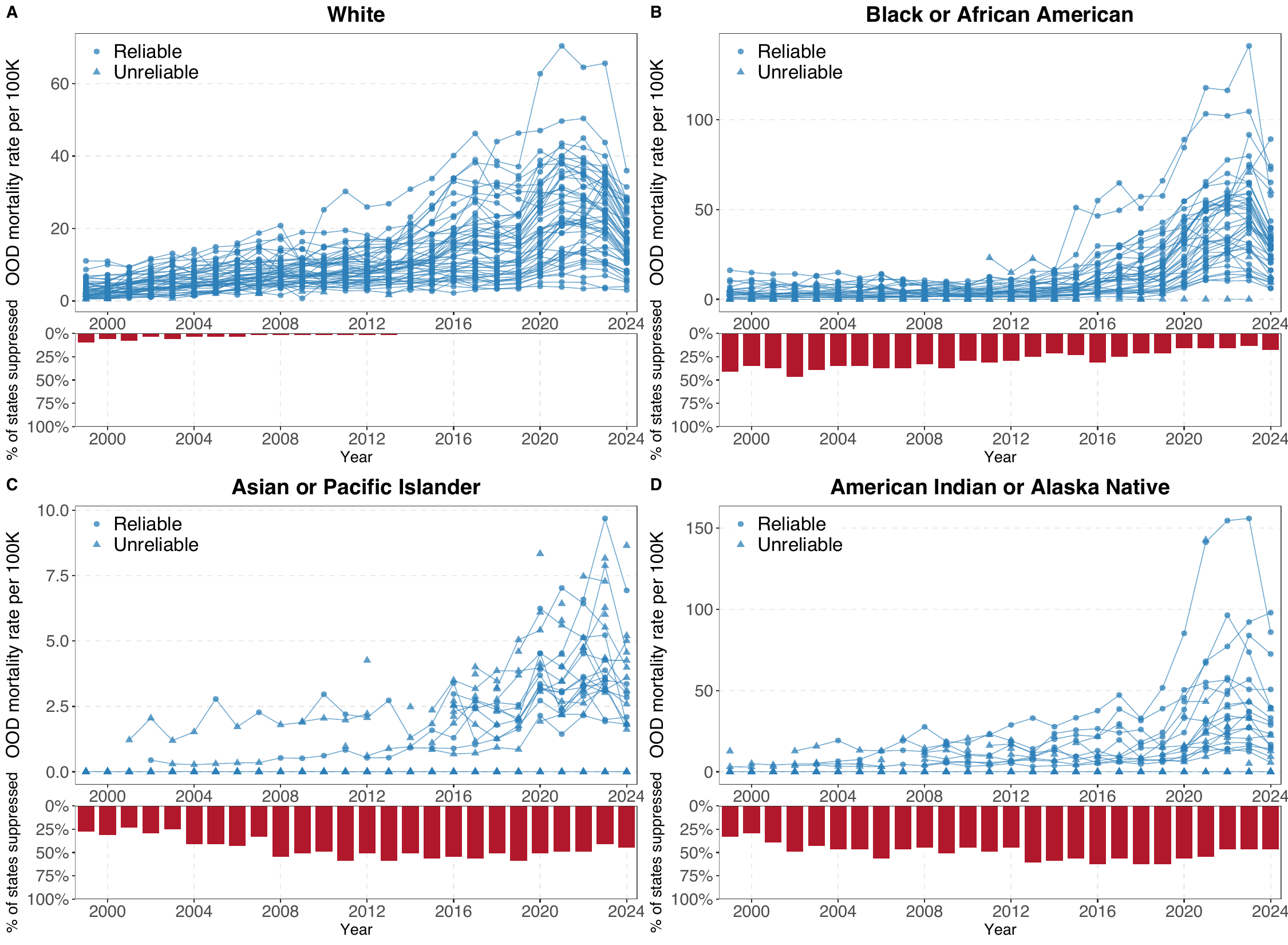}
    \caption{Temporal trends of state-level observed OOD mortality (per 100,000 population) and suppression rates (across the 50 U.S. states and the District of Columbia) by racial groups from 1999 to 2024: (A) White; (B) Black or African American; (C) Asian or Pacific Islander; and (D) American Indian or Alaska Native.}
    \label{fig:ood_rate_race}
\end{figure}

Figure~\ref{fig:ood_rate_race} displays annual OOD mortality rates from 1999 to 2024 across the 50 U.S. states and the District of Columbia (hereafter collectively referred to as states), stratified by racial group. For each racial group, the upper panel displays the mortality rate trajectory, while the lower panel shows the yearly percentage of states with suppressed counts, illustrating the extent of disclosure-induced suppression over time. Corresponding state-level OOD death counts are presented in Figure~S.1 of the Supplementary Materials. The observed epidemic trajectories vary substantially both across racial groups and among states. 

Rates generally increased after the mid-2010s, peaked around 2022--2023, and declined in 2024, although the timing, magnitude, and the shape of these temporal trajectories varied considerably across states and racial groups. Among states with unsuppressed data, median state-level rates peaked at approximately 24 per 100,000 for White populations, 46 for Black populations, 4 for Asian/PI populations, and 45 for AIAN populations. While the highest numbers of OOD deaths occurred among White populations, mortality rates in the later years of the epidemic were generally higher among Black and AIAN populations. By 2024, suppression had essentially disappeared for White populations and declined to approximately 18\% of states for Black populations, but persisted in near half of states for the Asian/PI and AIAN groups. These contrasting patterns illustrate that the populations with the greatest uncertainty due to data suppression often experience the highest mortality burden, highlighting the need for statistical methods that can infer subgroup-specific epidemic trajectories while appropriately accounting for disclosure-induced suppression. These challenges motivate the proposed Bayesian ACCESS model introduced in the next section, which directly models OOD death counts under the observed disclosure limitation mechanism, propagates suppression-induced uncertainty, and borrows information across states and racial populations while preserving meaningful subgroup differences.

\section{Bayesian Autoregressive Change-point and Clustering Estimation for Suppressed Count Series}
\label{sec:method}

In this section, we introduce the proposed Bayesian ACCESS (Autoregressive Change-point and Clustering Estimation for Suppressed Count Series) framework. Let $i=1,\ldots,I$ index geographic units, $k=1,\ldots,K$ index demographic groups, and $t=1,\ldots,T$ index time points. Let $y_{ikt}$ denote the underlying disclosure-free true count, and let $y^{(o)}_{ikt}$ denote the corresponding observed count, which may be subject to suppression. We let $n_{ikt}$ denote the population size and $\lambda_{ikt}$ denote the latent rate, interpreted as the expected count per unit population, for geographic unit $i$, demographic group $k$, and time point $t$. In the CDC WONDER OOD data described in Section~\ref{sec:data}, the geographic units are $I=51$ states, the demographic groups are $K=4$ racial groups, and the study period consists of $T=26$ years.

\subsection{Count Data Modeling and Suppression Mechanism}

Due to privacy protection requirements, the true count $y_{ikt}$ may only be observed through its suppressed counterpart $y^{(o)}_{ikt}$. Let $\mathcal{S}\subset \mathbb{N}$ denote the set of count values subject to suppression. Applying the suppression mechanism to the true count $y_{ikt}$ yields the observed value $y^{(o)}_{ikt}$:
\begin{equation*}
y^{(o)}_{ikt} = \mathrm{suppress}(y_{ikt})
=\begin{cases}
y_{ikt}, & \text{if } y_{ikt} \notin \mathcal{S},\\
\text{suppressed}, & \text{if } y_{ikt} \in \mathcal{S}.
\end{cases},\qquad \E (y_{ikt}\mid \lambda_{ikt})= \lambda_{ikt} n_{ikt},
\end{equation*}
where $\lambda_{ikt}\geq 0$ is the latent rate. Thus, when suppression occurs, the exact value of $y_{ikt}$ is unobserved but is known to belong to $\mathcal{S}$. For the CDC WONDER OOD data analyzed in this study, $\mathcal{S}=\{1,\ldots,9\}$.

Conditional on the latent rate $\lambda_{ikt}$, we assume that the distribution of latent count $y_{ikt}$ follows $F(n_{ikt}, \lambda_{ikt},\phi)$, where $F(n,\lambda,\phi)$ denotes a generic count distribution supported on $\mathbb{N}$, with parameter $n\in\mathbb{N}^+$, $\lambda> 0$ and additional parameter $\phi$. 
The Poisson distribution provides a natural starting point for modeling count data. However, in practice, counts often exhibit overdispersion, especially when data are aggregated across heterogeneous geographic units and demographic groups. To account for overdispersion, the Poisson observation model can be replaced by a negative binomial model with mean $n_{ikt}\lambda_{ikt}$ and overdispersion parameter $\phi>0$, providing additional flexibility and robustness to unobserved heterogeneity. In practice, the choice of $F$ can be guided by model selection criteria or predictive validation.

\subsection{Autoregressive Rate Model}

To characterize temporal dependence in the latent rates, we adopt the following autoregressive specification:
\begin{equation}\label{eq:ACCESS:autoregressive}
\begin{aligned}
\lambda_{ikt} &= \alpha_{ikt}+\beta_{ikt}\lambda_{ik(t-1)} \\
&=\eta_k\alpha_{it}+\zeta_k\beta_{it}\lambda_{ik(t-1)},
\end{aligned}
\end{equation}
where $\alpha_{ikt}\ge 0$ is the innovation parameter, representing the component of the current mortality rate not explained by the previous year's rate, and $\beta_{ikt}>0$ is the autoregressive dependence coefficient, which determines how strongly the previous year's rate is propagated into the current year. To separate shared geographic-temporal dynamics from demographic-group heterogeneity, we decompose these parameters into common geographic-temporal parameters $(\alpha_{it},\beta_{it})$ and demographic-group multiplicative effects $(\eta_k,\zeta_k)$. Specifically, $\alpha_{it}\ge 0$ and $\beta_{it}>0$ represent the innovation and autoregressive dependence parameters for unit $i$ in time point $t$, respectively. Parameters $\eta_k>0$ and $\zeta_k>0$ define the demographic-group multiplicative effects through $\alpha_{ikt}=\eta_k\alpha_{it}$ and $\beta_{ikt}=\zeta_k\beta_{it}$, capturing demographic heterogeneity in the innovation and autoregressive dependence components, respectively.

In contrast to commonly used log-linear Poisson regression models for count time series \citep[e.g.,][]{pedroza2006bayesian,dass2015clustering}, the proposed specification uses an identity-link autoregressive structure and models the rate directly as a function of its previous value. This formulation preserves a direct interpretation of the model parameters while allowing flexible temporal behavior. To illustrate, consider the simplified case in which $\alpha_{ikt}=\alpha$ and $\beta_{ikt}=\beta$ are constant over time. When $0<\beta<1$, the rate converges to the equilibrium level $\alpha/(1-\beta)$ geometrically as $t\to\infty$. When $\beta=1$, the rate follows a linear trend with increment $\alpha$ at each time step. When $\beta>1$, the rate exhibits exponential growth. Thus, the model can represent stable trajectories, linear growth, and rapid exponential trends within a single autoregressive framework. This specification can be viewed as an adaptation of integer-valued GARCH (INGARCH) models \citep{ferland2006integer} to multivariate suppressed count data with heterogeneous temporal dynamics. Closely related formulations have also been used in epidemic modeling \citep{held2005statistical}, where the intercept and autoregressive components are commonly interpreted as endemic and epidemic components, respectively. %Although our model shares a similar mathematical structure, its interpretation differs because mortality rates do not arise from a direct infectious-disease transmission process.

\subsection{Change-Point Modeling with Geographic Clustering}

To model abrupt changes in the temporal dynamics of each latent mortality trajectory, we assume the geographic-temporal autoregressive parameters are piecewise constant over time. 

Let $L_i\in\mathbb{N}$ denote the number of change points for unit $i$, and let $\vtau_i=(\tau_{i1},\ldots,\tau_{iL_i})$ denote the ordered change-point locations, with $1<\tau_{i1}<\cdots<\tau_{iL_i}<T$. Using the conventions $\tau_{i0}=0$ and $\tau_{i(L_i+1)}=T$, the parameters $(\alpha_{it},\beta_{it})$ are assumed to be constant within each segment $(\tau_{i(\ell-1)},\tau_{i\ell}]$, for $\ell=1,\ldots,L_i+1$. Thus, each time series is partitioned into $L_i+1$ contiguous temporal segments within which the autoregressive dynamics remain unchanged, while allowing structural shifts between segments.

Because counts observed at the geographic-demographic level may be sparse and subject to substantial small-count suppression, we introduce a latent clustering structure to borrow information across geographic units with similar temporal dynamics. Let $\boldsymbol{g}=(g_1,\ldots,g_I)$ denote the cluster assignments, where $g_i\in\{1,\ldots,G\}$ is the cluster label for geographic unit $i$, and $G\geq 1$ is the number of clusters. Both $G$ and $\boldsymbol{g}$ are treated as unknown parameters. Units assigned to the same cluster $g$ share a common change-point structure $\vtau^{(g)}=\bigl(\tau^{(g)}_1,\ldots,\tau^{(g)}_{L^{(g)}}\bigr)$, where $L^{(g)}$ is the number of change points, and common segment-specific autoregressive parameters $\{(\alpha^{(g)}_\ell,\beta^{(g)}_\ell)\}_{\ell=1}^{L^{(g)}+1}$. Specifically, for any geographic unit $i$ in cluster $g_i$, we assume $L_i = L^{(g_i)}$, $\vtau_i=\vtau^{(g_i)}$, and $(\alpha_{it},\beta_{it})=(\alpha^{(g_i)}_\ell,\beta^{(g_i)}_\ell)$ if $\tau^{(g_i)}_{\ell-1}<t\le \tau^{(g_i)}_{\ell}$. Therefore, the latent rate model in \eqref{eq:ACCESS:autoregressive} can be written compactly as
\begin{equation*}\label{eq:ACCESS:main}
\lambda_{ikt}
=
\sum_{\ell=1}^{L^{(g_i)}+1}
\mathbf{1}_{(\tau^{(g_i)}_{\ell-1}, \tau^{(g_i)}_{\ell}]}(t)
\left[
\eta_k\alpha^{(g_i)}_{\ell}
+
\zeta_k\beta^{(g_i)}_{\ell}\lambda_{ik(t-1)}
\right],
\end{equation*}
where $\mathbf{1}_{A}(\cdot)$ denotes the indicator function of the set $A$.
\subsection{Prior specification}

We complete the Bayesian specification by assigning priors to all model parameters, including the autoregressive parameters, the cluster configuration, and the change-point structure.

\subsubsection{Clustering prior}
We place a mixture of finite mixtures (MFM) prior \citep{miller2018mixture} on the unknown partition of the geographic units. Specifically, the prior on the cluster assignments $\vg = (g_1,\ldots,g_I)$ is modeled as
\begin{align*}
g_i \mid G, \boldsymbol{\pi} &\sim \mathrm{Categorical}(\pi_1,\dots,\pi_G),
\qquad i=1,\dots,I,\\
(\pi_1,\dots,\pi_G) \mid G &\sim \mathrm{Dirichlet}(\gamma,\ldots,\gamma),\quad 
G \sim p_G(\cdot),
\end{align*}
where $p_G$ is the prior distribution for the number of clusters $G$, supported on the positive integers,  $\vpi=(\pi_1,\dots,\pi_G)$ denotes the nonnegative  mixture weights satisfying $\sum_{g=1}^G\pi_g=1$, and $\gamma>0$ controls the prior entropy of $\boldsymbol{\pi}$. This MFM prior induces a random partition of the geographic units and allows the number of occupied clusters to be learned from the data. In this work, following \citet{miller2018mixture}, we specify $p_G$ by assigning a truncated Poisson prior, with $G-1 \sim \mathrm{Poisson}(\lambda_G)$ truncated to $\{0,1,\ldots,G_{\max}-1\}$, where $G_{\max}$ is a prespecified upper bound on the number of clusters.

\subsubsection{Change-point prior}
For each cluster $g$, we assign a truncated Poisson prior to the number of change points: $L^{(g)} \sim \mathrm{Poisson}(\lambda_{\mathrm{L}})$ truncated to $\{0,1,\dots,L_{\max}\}$,
where $L_{\max}\geq 1$ is a prespecified maximum number of change points. Conditional on $L^{(g)}$, we assign a uniform prior to the change-point locations $\vtau^{(g)}$ over all admissible configurations satisfying a minimum spacing constraint, following the literature \citep{dass2015clustering,cappello2023bayesian}, which avoids nearly coincident change points and improves identifiability. 
Specifically, we consider:
\begin{equation*}
p\bigl(\tau^{(g)}_1,\dots,\tau^{(g)}_{L^{(g)}} \mid L^{(g)}\bigr)
\propto
\mathbf{1}_{\mathcal{T}_{T}(L^{(g)},d_{\min})}(\vtau^{(g)}),
\end{equation*}
where
$\mathcal{T}_{T}(L,d_{\min})=\left\{
(\tau_1,\ldots,\tau_L):
1<\tau_{1}<\ldots <\tau_L<T,\ 
\tau_{\ell}-\tau_{\ell-1}\ge d_{\min},\ 
\ell=2,\ldots,L
\right\}$ and  $d_{\min}\geq 1$ is a prespecified minimum distance. 

\subsubsection{Segment-specific autoregressive parameters}
For each cluster $g$ and segment $\ell$, we assign half-Cauchy and lognormal priors to the positive innovation parameters and autoregressive dependence coefficients, respectively:
\begin{equation*}
\alpha^{(g)}_\ell \sim \mathrm{HalfCauchy}(\sigma_{\alpha}),
\qquad
\beta^{(g)}_\ell \sim \mathrm{LogNormal}(\mu_{\beta}, \sigma^2_{\beta}),
\end{equation*}
where $\sigma_{\alpha}>0$, $\mu_{\beta}\in\mathbb{R}$, and $\sigma^2_{\beta}>0$ are hyperparameters. The lognormal prior provides flexible support for positive autoregressive dependence coefficients while avoiding excessive prior mass near zero. In contrast, the half-Cauchy prior for the innovation parameters places substantial mass near zero, allowing regular temporal patterns to be largely explained by the lagged rate, while retaining heavy tails to accommodate large innovations.% when supported by the data.

\subsubsection{Demographic-group effects}

For the demographic-group effects, we impose symmetric identifiability constraints
$\prod_{k=1}^K\eta_k=\prod_{k=1}^K\zeta_k=1$ and assign constrained lognormal priors as follows. Let $\mH\in\R^{K\times(K-1)}$ be an orthonormal contrast matrix satisfying $\mH^\top\mH=\mI_{K-1}$ and $\mH^\top\mathbf{1}=\mb{0}$. We introduce latent log-contrast parameters with normal priors, $\boldsymbol{z}_{\eta}\sim \sN\left(\mb{0}_{K-1},(\sigma_\eta^2/2)\mI_{K-1}\right)$ and $\boldsymbol{z}_{\zeta}\sim \sN\left(\mb{0}_{K-1},(\sigma_\zeta^2/2)\mI_{K-1}\right)$, where $\sigma_\eta^2>0,\sigma_\zeta^2>0$ are hyperparameters. We then assign $\log\veta=(\log\eta_1,\ldots,\log\eta_K)^\top=\mH\boldsymbol{z}_{\eta}$ and $\log\vzeta=(\log\zeta_1,\ldots,\log\zeta_K)^\top=\mH\boldsymbol{z}_{\zeta}$. This construction enforces the product-one constraints without designating a reference demographic group. For any $k\neq k'$, the induced prior implies $\eta_k\sim\mathrm{LogNormal}(0,\sigma_\eta^2(K-1)/(2K))$ and $\eta_k/\eta_{k'}\sim\mathrm{LogNormal}(0,\sigma_\eta^2)$, with analogous results for $\zeta_k$ and $\zeta_k/\zeta_{k'}$. Thus, $\eta_k$ and $\zeta_k$ are interpreted as multiplicative effects %relative to the geometric mean 
across demographic groups, while pairwise ratios provide direct demographic-group comparisons. We provide a detailed proof of the prior properties for the demographic-group effects in Section~S.3.1 of the Supplementary Materials.
% }

\begin{figure}[ht!]
    \centering

    \includegraphics[width=\linewidth]{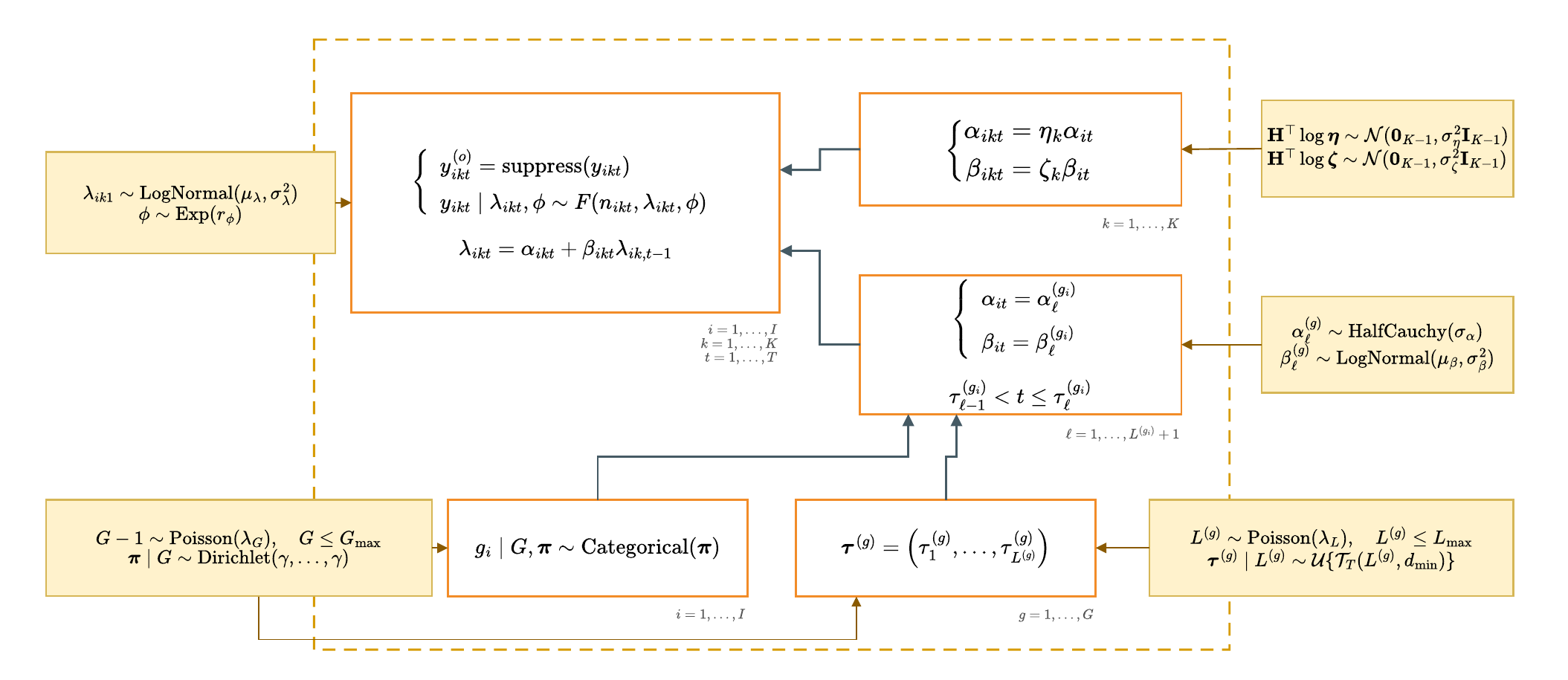}
    \caption{Illustration of the proposed Bayesian Autoregressive Change-point and Clustering Estimation for Suppressed Count Series (ACCESS).}
    \label{fig:bayesian_access_diagram}
\end{figure}

Finally, we assign independent log-normal priors $\lambda_{ik1}\sim\mathrm{LogNormal}(\mu_\lambda, \sigma^2_\lambda)$ for the initial rates, for $i\in\{1,\ldots,I\}$ and $k\in\{1,\ldots,K\}$. When the negative binomial observation model is used, we assign an exponential prior $\phi \sim \mathrm{Exp}(r_{\phi})$ with $r_{\phi}>0$ to place more mass near zero and encourage shrinkage toward the Poisson model while allowing overdispersion when supported by the data. The proposed Bayesian ACCESS model is illustrated in Figure~\ref{fig:bayesian_access_diagram}.

\subsection{Hyperparameter Specification for OOD Application}\label{subsec:hyperparam}

We adopt a weakly data-informed hyperparameter specification for the proposed model, which is used throughout the numerical studies. The hyperparameters are calibrated using prior scientific knowledge together with empirical scale information from the CDC WONDER OOD mortality data.

In the primary application, the data consist of $I=51$ states observed over $T=26$ years. We set $G_{\max}=10$ and $L_{\max}=4$ based on recent studies of state-level opioid-related policy and mortality trends \citep[e.g.,][]{jackson2020characterizing,bauer2024trends}. These choices provide sufficient flexibility for heterogeneous temporal dynamics while avoiding excessive model complexity and reducing computational burden. For the change-point prior, we set $\lambda_L = 1$ and impose a minimum spacing constraint of $d_{\min}=2$ %between consecutive change points 
to avoid unrealistically short temporal segments. For the clustering prior, we set $\gamma = 1$ and $\lambda_G = 4$, following a common default specification in the mixture-of-finite-mixtures literature.

It remains to specify the priors for the segment-specific autoregressive parameters and demographic-group effects. We adopt weakly informative priors with prior means chosen to reflect plausible parameter scales while retaining substantial prior uncertainty. Specifically, we set $\sigma_{\alpha} = 10^{-5}$ for the scale parameter of the Half-Cauchy prior of the autoregressive innovation terms. This scale is chosen to align with the observed national-level mortality rates. For the autoregressive dependence coefficients, we specify $(\mu_{\beta}, \sigma_{\beta}) = (0,0.5)$, which centers the prior around one while allowing moderate variability in temporal dependence. For the initial mortality rates, we set $(\mu_{\lambda}, \sigma_{\lambda}) = (-10, 2)$, corresponding to a prior median on the order of $10^{-5}$, reflecting the lower national mortality rates observed in the early 2000s. Finally, for the multiplicative demographic-group effects, we set $\sigma_{\eta} = \sigma_{\zeta} = 0.5$, yielding prior distributions  centered around $1$ while allowing substantial heterogeneity across racial groups. For the negative Binomial model, we set $r_\phi=40$ based on preliminary analysis, which penalizes large overdispersion values above $0.1$.

\subsection{Posterior Computation and Summary}\label{subsec:posterior:sampling_summary}
We implement the proposed Bayesian ACCESS model in \texttt{R} using NIMBLE \citep{de2017programming}, and obtain posterior samples of the model parameters via Markov chain Monte Carlo (MCMC). In practice, a naive implementation exhibits poor mixing, largely because the model involves a highly structured parameter space with both discrete variables, such as cluster assignments and change-point locations, and continuous parameters whose dimension varies with the clustering and segmentation structures. To address these computational challenges, we develop a tailored sampling scheme that enables more efficient exploration of the multimodal and highly structured posterior distribution. Specifically, for the clustering assignments, we use a collapsed Gibbs sampler by adapting Algorithm~8 of \cite{neal2000markov} to the MFM prior. For the change-point structures, we use categorical samplers over the admissible discrete support. For the remaining continuous parameters, we employ the adaptive block Metropolis algorithm \citep{roberts1997updating}. Finally, to explore the multimodal posterior distribution more efficiently, we adopt a parallel tempering scheme \citep[e.g.,][]{earl2005parallel}. We defer a more detailed description of the implementation and sampling algorithms to Section~S.3 of the Supplementary Materials.

We then introduce several summaries for interpreting the MCMC output. Let $\mathcal{D}$ denote the observed count and population data. For each geographic unit $i$, a direct quantity of interest is the posterior change-point probability $p_i(t\mid \mathcal{D})= P(t\in\{\tau_{i1},\ldots,\tau_{iL_i}\}\mid \mathcal{D})$, which measures the posterior evidence that a structural change occurs immediately after time $t$. Posterior inference for the cluster assignments $(g_1,\ldots,g_I)$ can be summarized by the posterior co-clustering matrix, whose $(i,j)$-th entry is $P(g_i=g_j\mid\mathcal{D})$, with larger values indicating greater similarity in temporal trends between geographic units. %A representative cluster configuration can then be selected using Dahl's least-squares method \citep{dahl2006model}. 
Posterior samples of the model parameters can also be used for prediction and aggregation while propagating uncertainty. Let $h\geq 1$ denote a forecasting horizon. Under the autoregressive model in \eqref{eq:ACCESS:autoregressive}, the posterior predictive distribution of the latent rate $\lambda_{ik(T+h)}$ can be obtained recursively as $\lambda_{ik(T+h)}=\alpha_{ikT}+\beta_{ikT}\lambda_{ik(T+h-1)}$, assuming no additional change point occurs after $T$. Finally, posterior samples can be aggregated across geographic units or demographic groups using population weights. For example, the geographically aggregated rate for demographic group $k$ at time $t$ is defined as $(\sum_{i=1}^I n_{ikt})^{-1}\sum_{i=1}^I n_{ikt}\lambda_{ikt}$. Similarly, we define the geographically aggregated posterior change-point probability as $(\sum_{i=1}^I n_{it})^{-1}\sum_{i=1}^I n_{it} p_i(t\mid \mathcal{D})$, where $n_{it}=\sum_{k=1}^K n_{ikt}$. Analogous population-weighted summaries can be constructed for other geographic or demographic aggregations of interest as well. Importantly, because prediction and aggregation are performed within each posterior sample, the resulting summaries incorporate uncertainty in the latent rates, autoregressive dynamics, and change-point structures, rather than conditioning on a fixed set of model parameters.

\section{Semi-synthetic Data Evaluation}\label{sec:simulation}
\subsection{Simulation Setup}

We conduct simulation studies to evaluate the performance of Bayesian ACCESS under varying suppression levels and temporal dynamics, using a realistic semi-synthetic setting calibrated from the CDC WONDER OOD mortality data. The goal is to assess whether the proposed model can recover latent rate trajectories, autoregressive parameters, cluster assignments, and change-point structures while providing reliable uncertainty quantification when only suppressed public-use counts are available. Additional details on parameter calibration, the data-generating process, and posterior summaries of model parameters are provided in Section~S.4.1 of the Supplementary Materials.

Following the OOD data described in Section~\ref{sec:data}, we consider $K=4$ demographic groups and $T=26$ time points. We construct four simulation settings by combining two geographic settings with two population-size settings. For geographic settings, rather than using the $51$ states directly, we consider two simplified settings based on the U.S. Census Bureau regional classification. In the \emph{Region} setting, the $51$ states are aggregated into $I=4$ U.S. Census regions with $G_0=2$ true clusters. In the \emph{Division} setting, they are aggregated into $I=9$ U.S. Census divisions with $G_0=3$ true clusters. Within each geographic setting, we vary population size to induce different degrees of data sparsity and suppression. In the \emph{Regular} setting, $n_{ikt}$ is defined as the sum of the corresponding state-level populations within geographic unit $i$, demographic group $k$, and year $t$. In the \emph{Thin} setting, the aggregated population is divided by $20$ to induce greater sparsity and suppression. Initial rates $\lambda_{ik1}$ are specified as population-weighted means of the observed initial state-level rates among the states within each geographic unit. Together, these choices yield four simulation settings with different clustering structures and degrees of data suppression.

For each cluster $g$ in each simulation setting, we specify a distinct change-point structure. When $g=2$, we set $L^{(g)}=2$ with $(\tau^{(g)}_1,\tau^{(g)}_2)=(10,20)$; otherwise, we set $L^{(g)}=1$ with $\tau^{(g)}_1=15$. Segment-specific autoregressive parameters $(\alpha^{(g)}_{\ell},\beta^{(g)}_{\ell})$ are calibrated from preliminary autoregressive estimates based on observed crude mortality rates from the CDC WONDER OOD data, with additional multiplicative perturbations introduced to induce heterogeneity across clusters and segments. The true demographic-group effects are specified as $(\eta_1,\eta_2,\eta_3,\eta_4)=(1,2,1/3,1.5)$ and $(\zeta_1,\zeta_2,\zeta_3,\zeta_4)=(1,0.9,10/9,1)$. %Given these parameters, latent rates $\lambda_{ikt}$ are generated recursively from the ACCESS autoregressive model. 
Under each simulation setting, we generate the model parameters and latent rate trajectories once, and then generate $10$ independent replicate datasets by sampling counts from the negative binomial observation model with mean $n_{ikt}\lambda_{ikt}$ and overdispersion parameter $\phi_0=0.05$.

We compare four methods. The first two are Bayesian ACCESS-based methods: the proposed Bayesian ACCESS model fitted to suppressed counts, denoted by ``ACCESS'', and an oracle version fitted to unsuppressed counts, denoted by ``Oracle ACCESS''. The remaining two methods are log-linear segmented regression approaches, fitted separately to each trajectory. The non-oracle version, denoted by ``Segmented'', is fitted to imputed complete data, with suppressed counts imputed using iterative proportional fitting (IPF). It reflects a common workflow in epidemiologic analyses involving change points and suppressed counts. The oracle version, denoted by ``Oracle Segmented'', is fitted to unsuppressed counts. Both segmented regression methods are implemented using the \texttt{segmented} package in \texttt{R}. 

We evaluate model performance using four metrics. For all four methods, latent rate estimation accuracy is assessed by the root mean squared error (RMSE),
$\mathrm{RMSE}_{\lambda}=\sqrt{(IKT)^{-1}\sum_{i=1}^I\sum_{k=1}^K\sum_{t=1}^T(\hat\lambda_{ikt}-\lambda_{ikt})^2}$,
where $\lambda_{ikt}$ is the true latent rate and $\hat\lambda_{ikt}$ is its estimate. Change-point detection accuracy is evaluated using the average Hausdorff distance (AHD) between estimated and true change-point sets, averaged over geographic units and posterior draws for ACCESS methods, and over geographic–demographic units for segmented regression methods. For ACCESS and Oracle ACCESS, we additionally assess recovery of the autoregressive parameters $\{\alpha_{ikt}\}$ and $\{\beta_{ikt}\}$ using $\mathrm{RMSE}_{\alpha}$ and $\mathrm{RMSE}_{\beta}$, defined analogously to $\mathrm{RMSE}_{\lambda}$. For the Bayesian methods (ACCESS and Oracle ACCESS), RMSEs are computed using posterior mean estimates.

\subsection{Simulation Results}

We fit Bayesian ACCESS using the default hyperparameter specification described in Section~\ref{subsec:hyperparam} and with $G_{\max}=I$. For each simulated dataset, we ran $3$ independent MCMC chains for $30{,}000$ iterations, discarded the first $15{,}000$ iterations as burn-in, and retained posterior samples every $5$ iterations thereafter, yielding  $9{,}000$ posterior samples in total.

We first summarize the degree of suppression induced under each simulation setting. Under the \emph{Regular} population setting, the average percentages of suppressed counts across $10$ replicates were $0.8\%$ for the \emph{Region} setting and $8.6\%$ for the \emph{Division} setting. Under the \emph{Thin} population setting, the corresponding suppression percentages increased to $37.6\%$ and $48.1\%$, respectively. Thus, the four settings cover a broad range of suppression levels, from nearly complete observation to substantial small-count suppression.

Table~\ref{tab:simulation_results} reports the performance metrics averaged over $10$ simulation replicates. Across all four settings, ACCESS consistently yields lower $\mathrm{RMSE}_{\lambda}$ and AHD than Segmented, indicating more accurate latent rate estimation and change-point detection. The improvements are particularly pronounced in the \emph{Division} setting with \emph{Thin} population, where suppression is substantial. These results highlight the benefits of explicitly modeling suppressed counts, borrowing information across related trajectories, and accommodating flexible autoregressive dynamics. Compared with Oracle ACCESS, ACCESS generally yields slightly higher RMSEs and AHDs, reflecting the information loss induced by suppression. Nevertheless, its performance remains comparable to that of the oracle model across all four settings with varying suppression levels. Moreover, both ACCESS and Oracle ACCESS recover the true clustering structure with high accuracy, achieving posterior expected adjusted Rand indices greater than $0.995$ across all settings. Overall, these results demonstrate that Bayesian ACCESS can accurately recover latent rate trajectories and change-point configurations from disclosure-limited data, even under substantial suppression. Finally, in Section~S.4.3 of the Supplementary Materials, we show that the MCMC chains generally mix well and the posterior distributions concentrate near the true parameter values, providing empirical support for the algorithm's performance and model identifiability.

\begin{table}[htbp]
\centering
\scriptsize
\setlength{\tabcolsep}{2pt}
\renewcommand{\arraystretch}{0.85}
\caption{Semi-synthetic evaluation of Bayesian ACCESS and competing methods over $10$ replicates. Entries are the mean with standard deviation in parentheses.}
\label{tab:simulation_results}
\begin{tabular}{l|l|l|c c c c}
\hline
Geographic setting & Population & Method & $10^5\mathrm{RMSE}_{\lambda}$ & AHD & $10^5\mathrm{RMSE}_{\alpha}$ & $\mathrm{RMSE}_{\beta}$ \\
\hline
\multirow{8}{*}{Region}
& \multirow{4}{*}{Regular} & ACCESS & 0.889 (0.309) & 0.172 (0.200) & 0.129 (0.0454) & 0.0318 (0.0132) \\
& & Oracle ACCESS & 0.882 (0.304) & 0.130 (0.0921) & 0.127 (0.0454) & 0.0310 (0.0134) \\
& & Segmented & 2.14 (0.523) & 3.48 (0.550) & -- & -- \\
& & Oracle Segmented & 2.14 (0.523) & 3.48 (0.550) & -- & -- \\
\cline{2-7}
& \multirow{4}{*}{Thin} & ACCESS & 2.04 (0.763) & 0.543 (0.319) & 0.212 (0.0369) & 0.0370 (0.0110) \\
& & Oracle ACCESS & 1.71 (0.622) & 0.443 (0.303) & 0.184 (0.0422) & 0.0382 (0.0139) \\
& & Segmented & 3.85 (0.530) & 3.89 (1.01) & -- & -- \\
& & Oracle Segmented & 4.30 (2.22) & 3.54 (0.862) & -- & -- \\
\hline
\multirow{8}{*}{Division}
& \multirow{4}{*}{Regular} & ACCESS & 0.812 (0.192) & 0.714 (0.327) & 0.113 (0.0484) & 0.0234 (0.00610) \\
& & Oracle ACCESS & 0.799 (0.161) & 0.544 (0.273) & 0.0990 (0.0361) & 0.0231 (0.00659) \\
& & Segmented & 2.39 (0.162) & 6.37 (0.766) & -- & -- \\
& & Oracle Segmented & 2.39 (0.149) & 6.35 (0.748) & -- & -- \\
\cline{2-7}
& \multirow{4}{*}{Thin} & ACCESS & 2.55 (0.829) & 1.84 (0.971) & 0.330 (0.140) & 0.0682 (0.0251) \\
& & Oracle ACCESS & 1.74 (0.282) & 1.54 (0.601) & 0.257 (0.0808) & 0.0564 (0.0162) \\
& & Segmented & 7.92 (1.17) & 15.5 (1.25) & -- & -- \\
& & Oracle Segmented & 5.69 (0.742) & 14.8 (1.72) & -- & -- \\
\hline
\end{tabular}
\end{table}
\section{Analysis of CDC WONDER Opioid-related Overdose Mortality Trajectories}\label{sec:real_data_analysis}

We fit Bayesian ACCESS to the OOD mortality data described in Section~\ref{sec:data}, using the hyperparameter specification in Section~\ref{subsec:hyperparam}. For each of the Poisson and negative binomial observation models, we ran an MCMC chain for $12{,}000$ iterations and discarded the first $4{,}000$ iterations as burn-in, yielding $8{,}000$ posterior samples per model. We compared the two models using the widely applicable information criterion (WAIC). %\citep[WAIC;][]{watanabe2013widely}. 
The negative binomial model yielded a substantially lower WAIC than the Poisson model. The posterior mean of the overdispersion parameter $\phi$ was $0.053$, with a $95\%$ credible interval (CrI) of $(0.049,0.057)$, indicating overdispersion in the observed counts. We therefore focus on the negative binomial model in the subsequent analysis. Detailed implementation procedures, posterior summaries of OOD mortality rates and model parameters, geographic and demographic aggregation results, and sensitivity analyses are provided in Section~S.5 of the Supplementary Materials.

\begin{figure}[ht!]
    \centering

    \includegraphics[width = \textwidth]{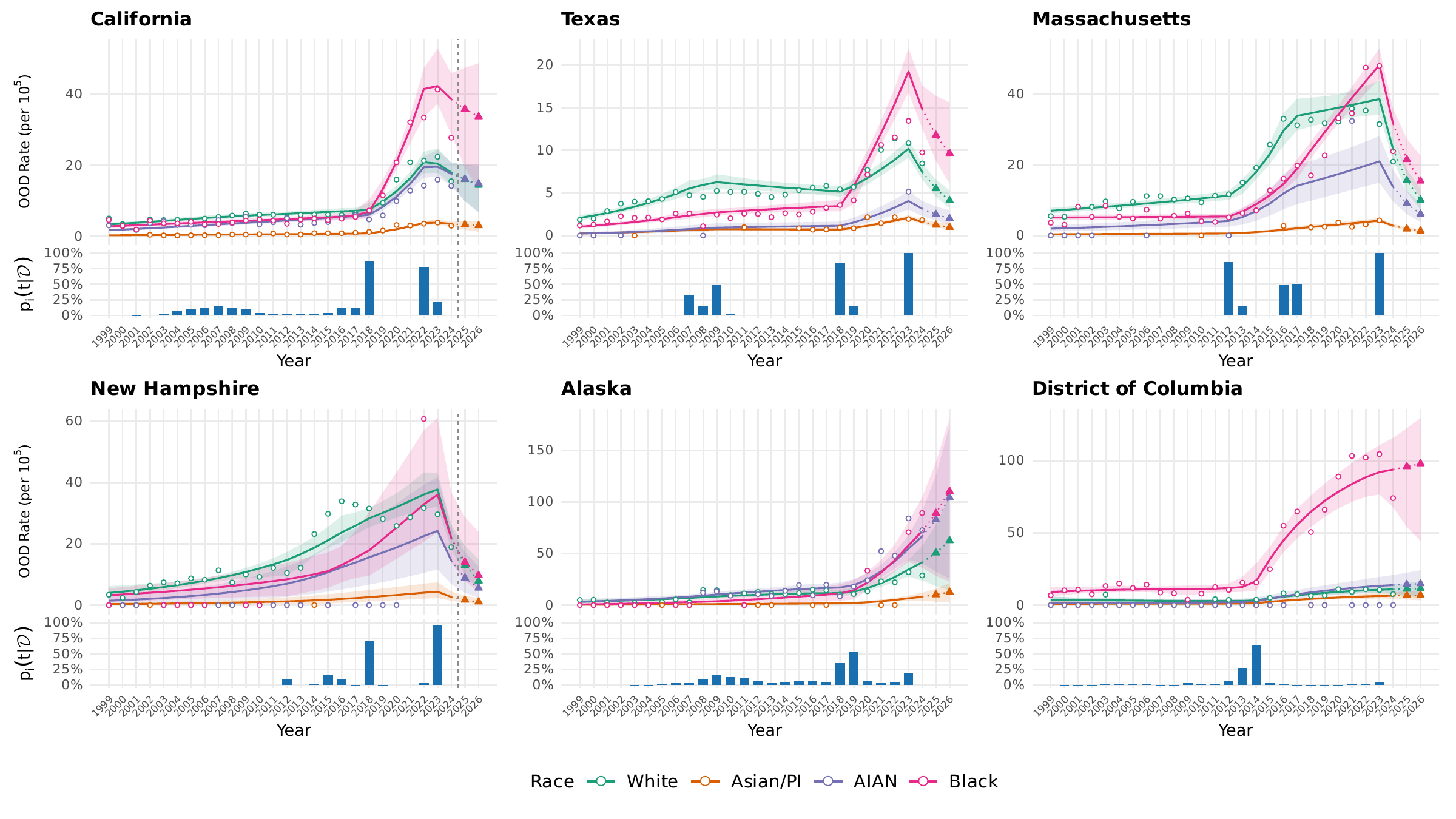}

    \caption{Posterior summaries of OOD mortality rates for six selected states. Dots denote observed mortality rates, with missing observations corresponding to CDC-suppressed years. Solid lines represent posterior mean mortality rates, shaded bands denote pointwise $95\%$ credible intervals, and triangles indicate posterior mean predictions for 2025 and 2026. The lower panel shows the posterior change-point probabilities $p_i(t\mid\mathcal{D})$.}
    \label{fig:selected_states_summary}
\end{figure}

We first present posterior summaries for six selected states in Figure~\ref{fig:selected_states_summary}. These states were chosen to illustrate heterogeneity in temporal mortality trajectories, change-point patterns, and degrees of suppression. California (CA), Texas (TX), and Massachusetts (MA) represent large-population states, whereas New Hampshire (NH), Alaska (AK), and the District of Columbia (DC) represent smaller or otherwise distinctive states with sparser counts and higher levels of suppression. For each state and racial group, we report the posterior mean mortality-rate trajectory with pointwise $95\%$ CrIs, posterior change-point probabilities, and posterior predictive mortality rates for 2025 and 2026. 

Across CA, TX, MA, and NH, we observe elevated posterior change-point probabilities near the end of the study period. In each state, the posterior change-point probability satisfies $p_i(t\mid\mathcal{D})\geq0.75$ in at least one year during 2022--2023, followed by a pronounced decline in the estimated mortality trajectory. This shared late-period shift is consistent with the sharp national decline in OOD mortality from 2023 to 2024, driven primarily by decreases in illicit fentanyl-involved deaths \citep{friedman2026charting,friedman2026us}. In contrast, AK and DC show weaker evidence of a recent structural change, with $p_i(t\mid\mathcal{D})\leq0.25$, and their posterior predictive trajectories retain considerable uncertainty regarding whether mortality will continue to decline after 2024. The selected states also exhibit heterogeneous evidence of earlier structural changes. CA, TX, MA, NH, and AK show elevated posterior change-point probabilities in the late 2010s, whereas MA, NH, and DC show elevated probabilities in the early to mid-2010s. These shifts are generally followed by steeper increases in mortality rates. Such structural changes are broadly consistent with the multi-wave U.S. opioid epidemic, including the transition from prescription opioids and heroin toward illicit fentanyl and other synthetic opioids after 2013, as well as the increasing involvement of fentanyl with stimulants in the mid-2010s \citep[e.g.,][]{ciccarone2019triple}. 
From a clustering perspective, Bayesian ACCESS identifies a representative partition with eight clusters using Dahl's method. The posterior co-clustering probabilities $P(g_i=g_j\mid\mathcal{D})$ indicate strong similarity in temporal dynamics among several groups of states. For example, the West Coast states CA, Oregon, and Washington have posterior co-clustering probabilities close to one. Nebraska, Nevada, and TX also exhibit high posterior co-clustering probabilities, suggesting similar OOD temporal patterns despite their geographic separation.
Overall, these summaries show that Bayesian ACCESS recovers both shared national-scale mortality temporal trend and substantial state-level heterogeneity. We note that the detected change points should be interpreted as statistical evidence of changes in mortality trajectories, not as causal effect estimates of specific policies or changes in drug-supply.

\begin{figure}[ht!]
    \centering
    \includegraphics[width=0.8\textwidth]{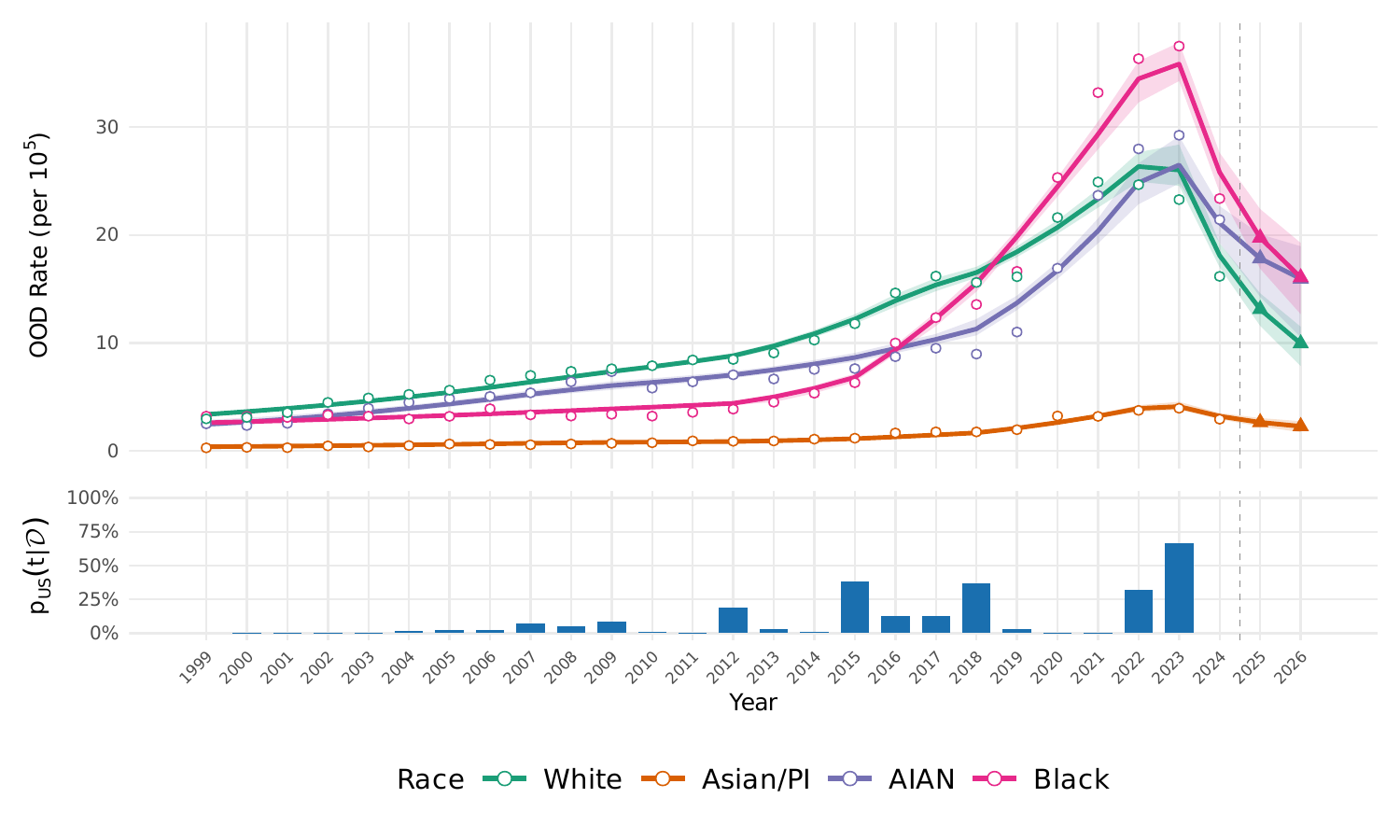}
    \caption{National OOD mortality rates by racial group. Dots denote observed national rates computed from national death counts and population. Solid lines represent posterior mean rates, shaded bands denote pointwise $95\%$ credible intervals, and triangles indicate posterior mean predictions for 2025 and 2026, all obtained through population-weighted aggregation. The lower panel shows the national posterior change-point probabilities $p_{\mathrm{US}}(t\mid\mathcal{D})$.}
    \label{fig:national_rate}
\end{figure}

Bayesian ACCESS produces posterior mortality-rate trajectories and predictions at multiple aggregation levels through population-weighted aggregation of the fitted state-race rates. We assess the national-level model fit by comparing aggregated posterior estimates with observed OOD mortality rates (Figure~\ref{fig:national_rate}). National counts are sufficiently large enough to calculate the observed rates directly from CDC WONDER death and population totals without suppression adjustment. %The model-based national rates, by contrast, are obtained by aggregating the posterior state-level rates. 
Their close agreement with the population-weighted posterior estimates indicates that the proposed model captures the overall mortality patterns well, supporting subsequent prediction and aggregation. The predicted percentage changes in 2026 relative to estimated rates in 2024 are summarized by racial group using posterior means and $95\%$ CrIs. The largest predicted decline is observed for White populations, at $-45.11\%$ ($-55.84\%$, $-39.45\%$), followed by Black populations, at $-37.89\%$ ($-47.83\%$, $-29.31\%$). Asian/PI and AIAN populations show comparatively smaller predicted declines of $-29.30\%$ ($-44.78\%$, $-20.44\%$) and $-24.66\%$ ($-46.80\%$, $-15.01\%$), respectively. Finally, the lower panel of Figure~\ref{fig:national_rate} reports national-level change-point probabilities $p_{\mathrm{US}}(t\mid\mathcal{D})$, obtained by population-weighted aggregation of state-level probability estimates. Years with elevated posterior evidence generally align with the evolving, multi-wave nature of the U.S. overdose crisis, including shifts in the dominant opioid types and changing patterns of substance involvement \citep{jalal2018changing,ciccarone2019triple,friedman2026charting,friedman2026us}.

\begin{figure}[ht!]
    \centering
    \includegraphics[width=\linewidth]{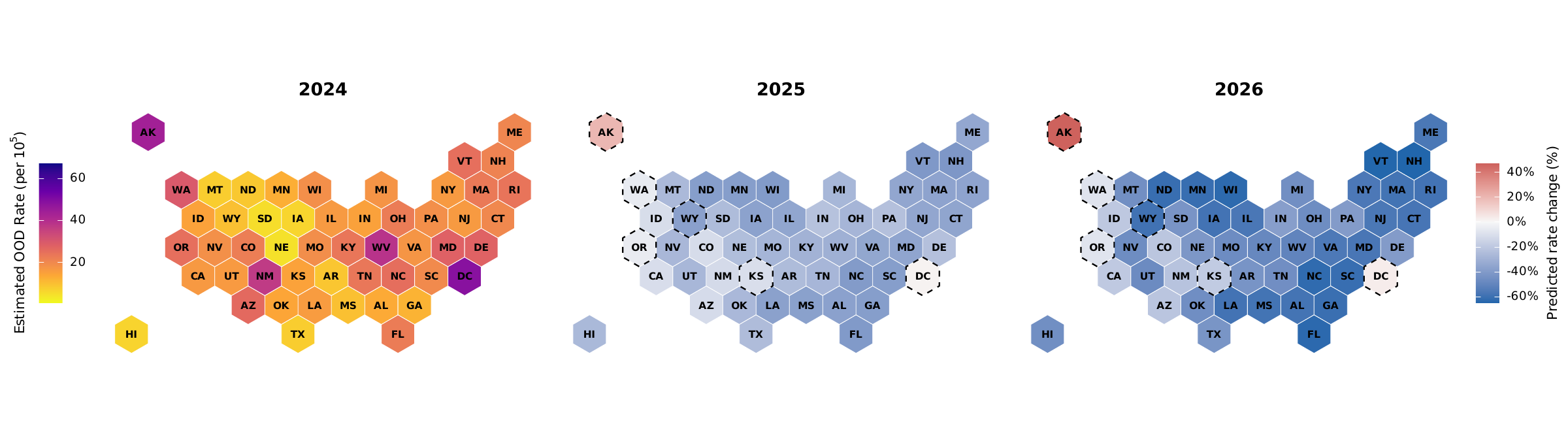}
    \caption{State-level race-aggregated OOD mortality rates and predicted changes. The first panel shows posterior mean mortality rates in 2024. The remaining panels show posterior mean predicted percentage changes in 2025 and 2026 relative to 2024, with dashed boundaries indicating states whose $95\%$ credible intervals include zero.}
    \label{fig:predicted_maps}
\end{figure}

We next examine geographic variation in predicted future OOD mortality using the fitted model. Following Section~\ref{subsec:posterior:sampling_summary}, we derive race-aggregated state-level mortality rates and compare posterior predictions for 2025 and 2026 with estimated rates in 2024. The left panel of Figure~\ref{fig:predicted_maps} shows the posterior mean state-level mortality rates in 2024. Nebraska and South Dakota have the lowest estimated OOD mortality rates, at approximately $5$ per $100{,}000$ population, with $97.5\%$ posterior quantiles remaining below $7$. In contrast, the District of Columbia, Alaska, and West Virginia have the highest estimated rates, at approximately $40$ per $100{,}000$ population, with $2.5\%$ posterior quantiles exceeding $25$. The two right panels show the posterior mean predicted percentage changes of mortality rates in 2025 and 2026 relative to 2024, with dashed boundaries indicating states whose $95\%$ CrIs include zero. The predictions suggest an overall decline in OOD mortality over the next two years, with substantial geographic variation in both magnitude and uncertainty. For 2026, $40$ states have posterior mean declines exceeding $35\%$ relative to 2024, with the upper bounds of their $95\%$ CrIs indicating declines of at least $25\%$, providing strong evidence of decreasing mortality. Several states show particularly large posterior mean declines, including New Hampshire ($-64.45\%$) and Wisconsin ($-62.48\%$). The remaining states show moderate declines or greater uncertainty. For example, the District of Columbia and Alaska have positive posterior mean percentage changes, while their $95\%$ CrIs are wide and include zero, indicating non-negligible uncertainty about whether mortality will decline. 

Finally, Table~\ref{tab:race_effects} summarizes the posterior estimates of the demographic-group effects, %Recall that the identifiability constraints $\prod_{k=1}^K\eta_k=\prod_{k=1}^K\zeta_k=1$ allow these multipliers to be interpreted through comparisons across racial groups. 
indicating racial heterogeneity in both components of the autoregressive rate model. For the innovation multiplier $\eta_k$, Asian/PI has the lowest estimate, followed by White and AIAN. Black has a substantially higher estimate than the other racial groups, indicating a markedly larger innovation component. For the autoregressive dependence multiplier $\zeta_k$, White and Asian/PI have similar estimates near one, while AIAN exhibits stronger and Black weaker autoregressive dependence. These results suggest that racial heterogeneity in OOD dynamics arises from distinct combinations of innovation and autoregressive dependence. By allowing both components to vary across racial groups, Bayesian ACCESS flexibly captures demographic-group differences in mortality trajectories, as illustrated in Figure~\ref{fig:selected_states_summary}.

\begin{table}[htbp]
\centering
\renewcommand{\arraystretch}{0.8}
\caption{Posterior summaries of demographic-group effects. The multiplicative effect $\eta_k$ acts on the innovation parameter $\alpha_{it}$, while $\zeta_k$ acts on the autoregressive dependence coefficient $\beta_{it}$. Estimates are posterior means with $95\%$ credible intervals in parentheses.}
\label{tab:race_effects}
\footnotesize
\begin{tabular}{l|l|c}
\hline
Effect & Racial group & Estimate \\
\hline
\multirow{4}{*}{$\eta_k$}
    & White & 0.605 (0.470, 0.742) \\
    & Asian or Pacific Islander & 0.375 (0.306, 0.462) \\
    & American Indian or Alaska Native & 0.727 (0.512, 0.932) \\
    & Black or African American & 6.219 (5.450, 7.108) \\
\hline
\multirow{4}{*}{$\zeta_k$}
    & White & 1.004 (0.999, 1.010) \\
    & Asian or Pacific Islander & 1.001 (0.994, 1.009) \\
    & American Indian or Alaska Native & 1.025 (1.017, 1.032) \\
    & Black or African American & 0.971 (0.963, 0.978) \\
\hline
\end{tabular}
\end{table}
\section{Discussion}\label{sec:discussion}

In this article, we introduced Bayesian ACCESS, a Bayesian hierarchical model for uncovering latent epidemic trajectories and identifying change points from publicly released suppressed count time series. The model characterizes temporal dynamics using an integer-valued autoregressive specification, enabling flexible and interpretable parameterization. Demographic heterogeneity is captured through multiplicative effects, while geographic heterogeneity is represented through a latent clustering structure. Through extensive numerical studies, we demonstrate that Bayesian ACCESS accurately captures the temporal dynamics of OOD mortality rates across U.S. states and racial groups, even under substantial data suppression, by efficiently sharing information across subpopulations. The proposed framework also provides change-point estimation with uncertainty quantification, allowing posterior evidence for structural changes to be directly assessed.

Despite its flexibility, the proposed model could be extended in several directions to better accommodate certain application-specific features. One direction is to enrich the autoregressive structure. Unlike classical INGARCH models \citep{ferland2006integer} and the endemic--epidemic formulation of \citet{held2005statistical}, our specification lets the latent rate $\lambda_{ikt}$ depend only on the previous latent rate $\lambda_{ik(t-1)}$, rather than also incorporating the lagged observed count $y_{ik(t-1)}$. It also uses a first-order autoregressive structure. Future extensions could include lagged counts, higher-order lag dependence, or cross-demographic temporal dependence, allowing $\lambda_{ikt}$ to depend on $\lambda_{ik'(t-h)}$ and $y_{ik'(t-h)}$ for $k\in\{1,\ldots,K\}$ and $h\geq1$. In the present analysis, we use the current first-order latent autoregressive specification because it avoids reliance on potentially unavailable lagged observed counts under data suppression and retains interpretable model parameters. Another possible extension concerns the clustering prior. A related alternative is the Dirichlet process prior \citep{teh2010dirichlet}; however, compared with the MFM framework, it tends to produce more small clusters and offers less direct control over the prior distribution on the number of clusters \citep{miller2018mixture}. Spatial dependence among geographic units could also be incorporated through adjacency-informed priors, such as a Markov random field constrained mixture of finite mixtures \citep[e.g.,][]{meng2026spatially}. In the present analysis, we do not impose such a constraint because adjacent states need not share similar OOD mortality trajectories: neighboring states may differ substantially in overdose-related policies, healthcare access, treatment infrastructure, and other contextual factors. %We therefore allow clustering to be driven by similarity in temporal dynamics rather than by geographic proximity.

Although motivated by OOD mortality, the proposed framework is broadly applicable to publicly released health statistics in which disclosure limitation results in systematically suppressed counts. Many federal and state health data systems routinely suppress small cell counts to protect confidentiality, creating similar inferential challenges for studies of infectious diseases, cancer incidence and mortality, maternal and child health, chronic disease surveillance, injury epidemiology, and other rare health outcomes. Rather than treating suppression as a preprocessing problem, the proposed framework directly incorporates the disclosure mechanism into a unified probabilistic model, enabling inference on latent spatiotemporal trajectories and their structural changes while appropriately propagating uncertainty arising from suppressed observations. More generally, this work illustrates how principled statistical modeling can expand the scientific utility of publicly released health statistics without requiring access to restricted individual-level data.

\bibliography{suppression_ref}

\end{document}